\documentstyle[11pt,csbmoriond,epsfig]{article}

\def\Journal#1#2#3#4{{#1} {\bf #2}, #3 (#4)}

\def\NPB{{\em Nucl. Phys.} B}
\def\PLB{{\em Phys. Lett.} B}
\def\PRL{\em Phys. Rev. Lett.}
\def\PRD{{\em Phys. Rev.} D}

\def\mco{\multicolumn}

\def\ra{\rightarrow}

\def\ko{K^0}

\def\be{\begin{equation}}
\def\ee{\end{equation}}
\def\bea{\begin{eqnarray}}
\def\eea{\end{eqnarray}}

\newcommand{\lesssim}{\stackrel{\scriptscriptstyle<}{\scriptscriptstyle\sim}}

\begin{document}

\rightline{UH-511-970-00}
\rightline{hep-ph/0008160}

\vspace*{4cm}
\title{HIGGS TRANSVERSE MOMENTUM AT THE LHC}

\author{ C. BAL\'AZS }

\address{Department of Physics and Astronomy, University of Hawaii,\\
Honolulu, HI 96822, U.S.A.}

\maketitle

\thispagestyle{empty}

\abstracts{
Resummed QCD corrections to the transverse momentum ($Q_T$) distribution 
of the Standard Model Higgs boson, produced at the CERN Large Hadron 
Collider, are presented. The small $Q_T$ factorization formalism is 
reviewed, which is used to extend the standard hadronic factorization 
theorem to the low $Q_T$ region, with emphasis on the matching to the 
standard hadronic factorization. Comparison of the $Q_T$ predictions from 
the extended factorization and the parton shower method is performed.
}

\section{Introduction}

One of the fundamental open problems of the Standard Model (SM) is 
revealing the dynamics of the electroweak symmetry breaking (EWSB). The 
physical remnant of the spontaneous EWSB, the Higgs boson, is the primary 
object of search at present and future colliders.\,\cite{HiggsSearch} 
At the 14 TeV center of mass energy Large Hadron Collider (LHC) at CERN, 
the SM Higgs boson will mainly be produced in proton--proton collisions 
through the partonic subprocess $g g$ (via top quark loop) $\to H 
X$.
To experimentally understand the Higgs signature, 
to enhance the statistical significance of the signal, and to measure its 
basic properties (mass, lifetime, spin, charge, couplings), it is necessary 
to determine the transverse momentum ($Q_T$) of the Higgs bosons. 

To reliably predict the $Q_T$ distribution of Higgs bosons at the LHC, 
especially for the low to medium $Q_T$ region where the bulk of the rate 
is, the effects of the multiple soft--gluon emission have to be included. 
This, and the need for the systematic inclusion of the higher order QCD 
corrections require the extension of the standard hadronic factorization 
theorem to the low $Q_T$ region. With a smooth matching to the usual 
factorization formalism, it is possible to obtain a prediction in the full 
$Q_T$ range. Since Monte Carlo event generators are heavily utilized in the
extraction of the Higgs signal, it is crucial to establish the reliability of 
their predictions. Thus, results of the parton shower formalism are also 
compared to those of the analytic calculation.

\section{Low $Q_T$ Factorization}


In this section the low transverse momentum factorization formalism and 
its matching to the usual factorization is described. When calculating 
fixed order QCD corrections to the cross section ${d\sigma}/{dQ dQ_T^2 
dy}$ for the inclusive process $pp \to H X$, the factorization theorem is 
invoked. The relevant variables $Q$, $Q_T$, and $y$, are the invariant 
mass, transverse momentum, and rapidity of the Higgs boson, respectively. 
The standard factorization
\be 
d\sigma = 
\sum_{j_1, j_2} \int \frac{d \xi_1}{\xi_1} \frac{d \xi_2}{\xi_2}
f_{j_1/h_1}(x_1,Q)\,
d{\hat \sigma}_{j_1j_2}\left(\frac{x_1}{\xi_1},\frac{x_2}{\xi_2}\right)\,
f_{j_2/h_2}(x_2,Q) ,
\ee
a convolution in the partonic momentum fractions $x_1$ and $x_2$, fails 
when $Q_T \ll Q$ as a result of multiple soft and soft+collinear 
emission of gluons from the initial state. The ratio of the two very 
different scales in the partonic cross section ${\hat \sigma}_{j_1j_2}$
(identified by the partonic indices $j_i$), 
produces large logarithms of the form $\ln(Q^2/Q_T^2)$ (being singular 
at $Q_T = 0$), which are not absorbed by the parton distribution functions 
$f_{j/h}$, unlike the ones originating from purely collinear parton 
emission. As a result, the Higgs $Q_T$ distribution calculated using the 
conventional hadronic factorization theorem is unphysical in the low $Q_T$ 
region.


To resolve the problem, the differential cross section is split into a part 
which contains all the contribution from the logarithmic terms ($W$), and 
into a regular term ($Y$): 
\be 
\frac{d\sigma}{dQ dQ_T^2 dy} = W(Q,Q_T,x_1,x_2) + Y(Q,Q_T,x_1,x_2) ,
\ee
Since $Y$ does not contain potentially large logs, it can be calculated 
using the usual factorization. The $W$ term has to be evaluated 
differently, keeping in mind that failure of the standard factorization 
occurs because it neglects the transverse motion of the incoming partons 
in the hard scattering. As it is 
proven~\cite{CollinsSoperYTerm,CollinsSoperSterman}, small $Q_T$ 
factorization gives the cross section as a convolution of transverse 
momentum distributions 
\be
W(Q,Q_T,x_1,x_2) = 
\sum_{j_1, j_2} \int d^2 \vec k_T \ 
{\tt C}_{j_1/h_1}(Q,\vec k_T,x_1)\,
{\tt H}_{j_1j_2}(Q,Q_T)\,
{\tt C}_{j_2/h_2}(Q,\vec Q_T - \vec k_T,x_2) .
\ee
Here ${\tt H}_{j_1j_2}$ is a hard scattering function, and ${\tt C}_{j/h}$ 
are partonic density distributions depending on both longitudinal ($x$) 
and transverse ($k_T$) momenta, and on the scale of the factorization 
which is set equal to the hard scale $Q$. The convolution simplifies to a 
product in the Fourier conjugate, i.e. transverse position (${\vec b}$) space
\be
\widetilde{W}(Q,b,x_1,x_2) = 
{\cal C}_{j_1/h_1}(Q,b,x_1)\,
{\cal H}_{j_1j_2}(Q,b)\,
{\cal C}_{j_2/h_2}(Q,b,x_2) ,
\ee
where $\widetilde{W}$, ${\cal C}_{j/h}$ and ${\cal H}_{j_1j_2}$ are the 
Fourier transforms of $W$, ${\tt C}_{j/h}$ and ${\tt H}_{j_1j_2}$. The 
generalized parton distributions ${\cal C}_{j/h}$, together with the hard 
scattering function, satisfy an evolution equation somewhat similar to the 
usual DGLAP equations. 


The evolution equation, for the production of a colorless boson, takes the 
form\,\cite{CollinsSoperSterman}
\be
\frac{\partial}{\partial \ln Q^2}\widetilde{W}(Q,b,x_1,x_2) = 
- \int_{C_1/b^2}^{C_2 Q^2} \frac{d \mu^2}{\mu^2} 
\left[ 
A \left(\alpha_S(\mu),C_1\right) + 
B \left(\alpha_S(Q),C_1,C_2\right)
\right] ,
\ee
where $\alpha_S$ is the strong coupling constant.
It is customary to choose the renormalization scales arising in the 
evolution equation such that $C_1=2e^{- \gamma _E}\equiv C_0$ and $C_2=1$. 
The solution of the above evolution equation leads to the 
expression\,\cite{CollinsSoperSterman}
\be
\widetilde{W}(Q,b,x_1,x_2) = 
{\cal C}_{j_1/h_1}(Q,b,x_1)\,
e^{-{\cal S}(Q,b_*)}\,
{\cal C}_{j_2/h_2}(Q,b,x_2) ,
\ee
with the Sudakov exponent
\begin{equation}
{\cal S}(Q,b_*) = 
\int_{C_0^2/b_*^2}^{Q^2} \frac{d \mu^2}{\mu^2}
\left[
A \left( \alpha_S(\mu) \right) \ln \left( \frac{Q^2}{\mu^2} \right) + 
B \left( \alpha_S(\mu) \right)
\right] ,
\label{Eq:PerturbativeSudakov}
\end{equation}
which resums the large logarithmic terms.\,\footnote{To prevent evaluation 
of the Sudakov exponent in the non--perturbative region, the impact 
parameter $b$ was replaced by $b_* = b/\sqrt{1+(b/b_{\rm max })^2}$.} The 
partonic recoil against soft gluons as well as the intrinsic partonic 
transverse momentum are included in the modified parton distributions
\begin{equation}
{\cal C}_{j/h}(Q,b,x) =
\sum_a \left[ \int_x^1{\frac{d\xi }\xi }
C_{ja}\left( b_*,\frac{x}{\xi},Q \right) f_{a/h}(\xi,Q) \right]
{\cal F}_{a/h}(x,b,Q) .
\label{CalC}
\end{equation}
The $A$ and $B$ functions, and the Wilson coefficients $C_{ja}$ are free 
of logs and safely calculable perturbatively as expansions in the strong
coupling
\begin{equation}
A(\alpha_S) =
\sum_{n=1}^\infty 
\left( \frac{\alpha _S}\pi \right)^n A^{(n)}, ~~~ {\rm etc.}
\end{equation}
The process independent non--perturbative functions ${\cal F}_{a/h}$, 
describing long distance transverse physics, are extracted from low--energy 
experiments.\,\cite{Brock}


In the large $Q_T$ region, where $Q_T \sim Q$, the standard factorization 
theorem is applicable. The matching of the small $Q_T$ region to the large 
$Q_T$ result is achieved via the $Y$ piece. To correct the behavior of the 
resummed piece in the intermediate and high $Q_T$ regions, it is defined 
as the difference of the differential cross section calculated from the 
standard factorization formula at a fixed order $n$ of perturbation theory 
and its $Q_T\ll Q$ asymptote:\,\footnote{The expression of the $Y$ term 
for Higgs production can be found elsewhere.\,\cite{Yuan}}
\begin{equation}
Y(Q,Q_T,x_1,x_2)=
      \frac{d\sigma^{(n)}}{dQ^2\,dQ_T^2dy} -
\left.\frac{d\sigma^{(n)}}{dQ^2\,dQ_T^2dy}\right|_{Q_T\ll Q} .  
\label{Y.def}
\end{equation}
Using this definition, the cross section to order $\alpha_S^n$ is written as
\begin{equation}
\frac{d\sigma }{dQ^2\,dQ_T^2dy}=
W(Q,Q_T,x_1,x_2) +
      \frac{d\sigma^{(n)}}{dQ^2\,dQ_T^2dy} -
\left.\frac{d\sigma^{(n)}}{dQ^2\,dQ_T^2dy}\right|_{Q_T\ll Q} .  
\label{Eq:CSSMatched}
\end{equation}
At low $Q_T$, when the logarithms are large, the asymptotic part dominates 
the $Q_T$ distribution, and the last two terms cancel in 
Eq.~(\ref{Eq:CSSMatched}), leaving $W$ well approximating the cross section. 
At $Q_T$ values comparable to $Q$ the logarithms are small, and the expansion 
of the resummed term cancels the logarithmic terms up to higher orders in 
$\alpha_S$.\footnote{The cancellation is higher order than the order at 
which the singular pieces were calculated.} In this situation the first 
and third terms nearly cancel and the cross section reduces to the fixed order 
perturbative result. After matching the resummed and fixed order cross 
sections in such a manner, it is expected that the normalization of the 
cross section (\ref{Eq:CSSMatched}) reproduces the fixed order total rate, 
since when expanded and integrated over $Q_T$ it deviates from the fixed order 
cross section $\sigma^{(n)}$ only in higher order terms. 
Further details of the low $Q_T$ factorization formalism and its application to 
Higgs production can be found in the recent 
literature.\,\cite{BalazsYuanWZthesis}

\section{Higgs $Q_T$ at the LHC}

The low $Q_T$ factorization formalism, described in the previous section, 
is utilized to calculate the QCD corrections to the production of Higgs 
bosons at the LHC. 
In the low $Q_T$ region this 
calculation takes into account the effects of the multiple--soft gluon 
emission including the Sudakov exponent ${\cal S}$ and the non--perturbative 
contributions ${\cal F}_{a/h}$. In the Sudakov exponent the 
$A^{(1)}$, $A^{(2)}$, and $B^{(1)}$ coefficients are included. The ${\cal 
O}(\alpha_S^3)$ virtual corrections are also taken into account by 
including the Wilson coefficient ${\cal C}_{gg}^{(1)}$, which ensures the 
${\cal O}(\alpha_S^3)$ total rate. By matching to the ${\cal 
O}(\alpha_S^3)$ fixed order distributions a prediction is obtained for the 
Higgs production cross section in the full $Q_T$ range which is valid up 
to ${\cal O}(\alpha_S^3)$. The details of this calculation are given in an 
earlier work.~\cite{BalazsYuanH} The analytic results are coded in the 
ResBos Monte Carlo event generator.~\cite{BalazsYuanWZthesis}

\begin{figure}
\begin{center}
\epsfig{file=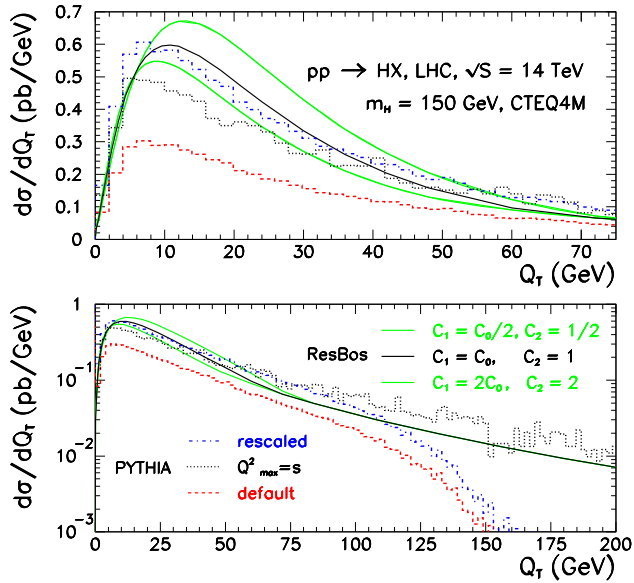,width=8.65cm}
\end{center}
\caption{
Higgs boson transverse momentum distributions calculated by ResBos 
(curves) and PYTHIA (histograms). The default (middle) ResBos curve was 
calculated with the canonical choice of the renormalization constants, and 
the other two with doubled (lower curve) and halved (upper curve) values 
of $C_1$ and $C_2$. For PYTHIA, the original output with default input 
parameters (dashed), the same rescaled by a factor of $K = 2$ (dash-dotted), 
and a curve calculated by the altered input parameter value $Q_{max}^2 = 
s$ (dotted) are shown. The lower portion, with a logarithmic scale, 
also shows the high $Q_T$ region.
\label{Fig:PYTHIA}}
\end{figure}

Fig.~\ref{Fig:PYTHIA} compares the Higgs boson transverse momentum 
distributions calculated by ResBos (curves) and by PYTHIA \cite{PYTHIA} 
(histograms from version 6.122). The middle solid curve is calculated 
using the canonical choice for the renormalization constants in the 
Sudakov exponent: $C_1 = C_0$, and $C_2 = 1$. To estimate the size of the 
uncalculated $B^{(2)}$ term, these renormalization constants are varied. 
The upper solid curve shows the result for $C_1 = C_0/2, C_2 = 1/2$, and 
the lower solid curve for $C_1 = 2C_0, C_2 = 2$. The band between these 
two curves gives the order of the uncertainty originating from the exclusion 
of $B^{(2)}$. The typical size of this uncertainty, e.g. around the peak 
region, is in the order of $\pm 10$ percent. The corresponding uncertainty 
in the total cross section is also in the same order. This uncertainty is 
larger than the uncertainty arising from the non--perturbative sector of 
the formalism (which was estimated to be less than 5 percent in the 
relevant $Q_T$ region).

\section{Comparison to Parton Showers}

Multiple soft--gluon radiation from the initial state can also be treated 
by the parton shower technique.\,\cite{PYTHIA} This approach is based on 
the usual factorization theorem, giving the probability ${\cal P} = 
e^{-{\cal S}(Q)}$ of the evolution from the scale $Q_0$ to $Q$, with no 
resolvable branchings by the exponent
\bea
{\cal S}(Q) =  
\int_{Q_0^2}^{Q^2} \frac{d \mu^2}{\mu^2} \frac{\alpha_S(\mu)}{2\pi} 
\int_{0}^{1} dz \, P_{a\to bc}(z),
\label{eq:ShowerS}
\eea
which is defined in terms of the DGLAP splitting kernels $P_{a\to bc}(z)$.
The formalism can be extended to soft emissions as well by using angular 
ordering. The distinct difference between the Sudakov exponents of the low 
$Q_T$ factorization and the parton showering approach is apparent. 
A comparison can be made based on the qualitative argument that parton 
showering resums leading logs which depend only on the given initial 
state. These logs correspond to the logs weighted by the $A$ function in 
Eq.(\ref{Eq:PerturbativeSudakov}). 

Fig.~\ref{Fig:PYTHIA} shows that the shape of the PYTHIA histogram agrees 
reasonably with the resummed curve in the low and intermediate $Q_T$ 
($\lesssim 125$ GeV) region. For large $Q_T$, the PYTHIA prediction falls 
under the ResBos curve, since ResBos mostly uses the exact fixed order 
${\cal O}(\alpha_S^3)$ matrix elements in that region, while PYTHIA still 
relies on the multi--parton radiation ansatz. PYTHIA can be tuned to agree 
with ResBos in the high $Q_T$ region, by changing the maximal virtuality a 
parton can acquire in the course of the shower (dotted curve). In 
that case, however, the low $Q_T$ region will have disagreement. 

Since showering is attached to a process after the hard scattering takes 
place, and the parton shower occurs with unit probability, it does not 
change the total cross section for Higgs boson production given by the 
hard scattering rate.
Thus, the total rate is given by PYTHIA at ${\cal O}(\alpha_S^2)$. In 
Fig.~\ref{Fig:PYTHIA} the dashed PYTHIA histogram is plotted without 
altering its output. For easier comparison the default PYTHIA histogram is 
also plotted after the rate is multiplied by the factor $K = 2$ (dash-dotted).
A detailed comparison of the results for Higgs boson production from 
ResBos and from event generators based on the parton shower algorithm is 
the subject of a separate work.\,\cite{BalazsHustonPuljak}

\section*{Acknowledgments}

I thank the organizers of the XXXVth Rencontres de Moriond for their 
hospitality and financial support. I am indebted to my collaborators in 
this work: J.C. Collins, J. Huston, I. Puljak, D.E. Soper, and C.--P. Yuan 
for many invaluable discussions. This work was supported in part by the 
DOE under grant DE-FG-03-94ER40833.

\end{document}